\documentstyle{article}
\title{The mathematical role of time and space-time in classical physics}
\author{Newton C. A. da Costa\\Research Group on Logic and Foundations.\\Institute for Advanced Studies, University of S\~{a}o Paulo.\\Av. Prof. Luciano Gualberto, trav. J, 374.\\05655-010 S\~{a}o Paulo SP Brazil. \and Adonai S. Sant'Anna\thanks{To whom correspondence should be sent. E-mail: adonai@scientist.com URL: http://www.geocities.com/adonaisantanna/adonaiss.html}\\Department of Mathematics, Federal University of Paran\'a\\C. P.
019081, 81531-990 Curitiba, PR, Brazil.}
\begin{document}
\maketitle
\newtheorem{definicao}{Definition}
\newtheorem{teorema}{Theorem}
\newtheorem{lema}{Lemma}
\newtheorem{corolario}{Corolary}
\newtheorem{proposicao}{Proposition}
\newtheorem{axioma}{Axiom}
\newtheorem{observacao}{Observation}

\begin{abstract}
We use Padoa's principle of independence of primitive symbols in axiomatic systems in order to discuss the mathematical role of time and space-time in some classical physical theories. We show that time is eliminable in Newtonian mechanics and that space-time is also dispensable in Hamiltonian mechanics, Maxwell's electromagnetic theory, the Dirac electron, classical gauge fields, and general relativity.
\end{abstract}


\section{Introduction}

	In a series of papers G. Jaroszkiewicz and K. Norton have developed a quite interesting research program where the continuum space-time is replaced by a discrete space-time in classical and quantum particle mechanics as well as in classical and quantum field theories \cite{Norton-97a,Norton-97b,Norton-98a,Norton-98b}. Many other authors have developed similar ideas \cite{Bender-85,Lee-83,Khorrami-95} in the last years. One of us has recently proposed the use of category theory in order to guarantee the consistency between discrete space-time and the continuum space-time picture for classical particle mechanics \cite{Sant'Anna-99}. Nevertheless, one of the pioneers in the discretization of space-time was the Japanese physicist T. Tati \cite{Tati-64,Tati-83}. Tati began his scientific career working with S. Tomonaga. Both of them were concerned with the divergences in quantum field theory. Tomonaga followed the way of renormalization theory. Tati tried the elimination of space-time in quantum theory in order to develop a theory of finite degree of freedom and avoid the divergences. Tomonaga won the Nobel prize. Tati's ideas about the elimination (by means of discretization) of space-time in physics were forgotten. If the reader takes a look on {\em Web of Science\/}, it will be easy to see that Tati's papers about non-space-time physics have no citation at all.

	This research program of discretization confirms two things: (i) space-time effectivelly plays a very fundamental role in physics, at least from the intuitive point of view; and (ii) discretization is a deep way to change the usual concepts about space-time. Discrete space-time seems to be more appropriate from the physical point of view than continuum space-time, at least in some cases. In a recent paper, e.g., C. Vafa \cite{Vafa-00} says:

\begin{quote}
Many ideas in physics sound like basic objects in number theory. For example, electrical charge comes in quanta, and do not take continuous values; matter comes in quanta, called particles, and do not come in continuous values; etc. In some sense quantum theory is a bending of physics towards number theory. However, deep facts of number theory play no role in questions of quantum mechanics. In fact it is very surprising, in my view, that this area of mathematics has found very little application in physics so far. In particular we do not know of any fundamental physical theories that are based on deep facts in number theory.
\end{quote}

	Yet, as a scientific prophet, Vafa predicts that in this century

\begin{quote}
...{\em we will witness deep applications of number theory in fundamental physics.\/}
\end{quote}

	W. M. Stuckey presents another perspective \cite{Stuckey-96,Stuckey-99,Stuckey-??}. According to him, the usual notion of differentiable manifold as a model of space-time is not appropriate in quantum physics. He does not appeal to any discretization program. He proposes a ``pregeometric reduction of transtemporal objects'' in order to cope with non-locality and quantum gravity.

	On the other hand, there is an insightful paper by U. Mohrhoff \cite{Mohrhoff-00} that presents a novel interpretation of quantum mechanics, where objective probabilities are assigned to counterfactuals and are calculated on the basis of all relevant facts, including those that are still in the {\em future\/}. According to this proposal, the intuitive distinction between here and there, past and future, has nothing to do with any physical reality:

\begin{quote}
The world is built on facts, and its spatiotemporal properties are supervenient on the facts. To understand how EPR [Einstein-Podolsky-Rosen {\em Gedanken\/} experiment] correlations are possible, one needs to understand that, in and of itself, physical space is undifferentiated. At a fundamental level, ``here'' and ``there'' are the same place. If EPR correlations require a medium, this identity is the medium.
\end{quote}

	Mohrhoff's ideas about space-time may inspire us for a new perspective about space-time, quite different from discretization. Tati believed that discretization of space-time is a manner to {\em eliminate\/} it from physics. But in this paper we discuss a different point of view, more radical, in a sense.

	We note, in a very general framework, that time and space-time are dispensable in some of the main classical physical theories. This corroborates the intuitive idea given above that space-time properties are supervenient on physical facts.

	Our results are simple, almost trivial, from the mathematical point of view, but, as we shall show in future papers, they are philosophically relevant. 

\section{Axiomatization of Physical Theories}

	The presence of mathematics in physics is grounded on the belief that there are mathematical patterns lurking in the dynamics of physical phenomena and natural laws. This is one of the most fundamental, powerful and `holy' beliefs in science \cite{Stewart-95}. On the other hand, mathematical concepts cannot, in principle, be considered as a faithful picture of the physical world. There is no length measurement or elapsed time interval, e.g., that actually corresponds to a real number, since measurements do have serious limitations, among others, of precision. This is one justification for the discretization of space-time in physics, according to our discussion in the Introduction. Nevertheless, the hypothesis of the continuum space-time is largely used in physical theories. Although mathematics is a science conceived by the human mind of  mathematicians, it has been used to send men to the Moon, to make computers, and to facilitate our lives.

	The axiomatic method is the soul of mathematics, the synthesis of the scientific method, and a powerful tool for the scientific philosopher. Rationality, in formal sciences, consists, in particular, in the correct, implicit or explict, use of the axiomatic method. 

	Roughly speaking, the axiomatic method begins with the definition of an appropriate language. In this language we have (i) the vocabulary, which is a set of symbols, refered to as primitive symbols; (ii) the grammar rules, usually refered to as an effective procedure to distinguish the {\em well-formed formulas\/} from other arbitrary sequences of symbols; (iii) a specific set of well-formed formulas called {\em axioms\/}; and (iv) some {\em rules of inference\/} which allow us to prove theorems from the axioms. This is possible because the rules of inference establish the relations of logical consequence among well-formed formulas.

	Primitive concepts are necessary since we cannot define everything. 

	The reason for the very notion of well-formed formulas is to distinguish between meaningful and meaningless expressions. The main purpose of the axioms is to settle up how the primitive concepts are related to each other in a list of assumptions, in order to derive other sentences and bits of information about these primitive concepts by means of the inference rules.

	It is not clear, from the historical point of view, if the axiomatic method was born as a philosophical or as mathematical subject. But it was used for the first time in ancient Greece, mainly represented by the famous book {\em Elements\/} by Euclide.

	One of the main goals of the axiomatic method in physics is the investigation of the logical foundations of physical theories. Following ideas delineated by Hilbert (1900) in his famous {\em Matematische Probleme}, a basic problem may be described as follows \cite{Hilbert-00}:

\begin{quote}
The investigations of the foundations of geometry suggest the problem: {\em To treat in the same manner, by means of axioms, those physical sciences in which mathematics plays an important part: first of all, the theory of probability and mechanics.\/}

\noindent
...

If geommetry is to serve as a model for the treatment of physical axioms, we shall try first by a small number of axioms to include as large a class as of physical phenomena, and then by adjoining new axioms to arrive gradually at the more special theories. At the same time Lie's principle of subdivision can perhaps be derived from profound theory of infinite transformation groups. The mathematician will have also to take account not only of those theories coming near to reality, but also, as in geometry, of all logically possible theories. He must be always alert to obtain a complete survey of all conclusions derivable from the system of axioms assumed.

Further, the mathematician has the duty to test exactly in each instance whether the new axioms are compatible with the previous ones. The physicist, as his theories develop, often finds himself forced by the results of his experiments to make new hypotheses, while he depends, with respect to the compatibility of the new hypotheses with the old axioms, solely upon these experiments or upon a certain physical intuition, a practice which in the rigorously logical building up of a theory is not admissible. The desired proof of the compatibility of all assumptions seems to me also of importance, because the effort to obtain each proof always forces us most effectually to an exact formulation of the axioms.
\end{quote}

	We believe that this kind of investigation may drive us to a better understanding of the actual role of fundamental concepts, and even fundamental principles, in mechanics and physics in general.

	In Euclidian geometry, e.g., it is possible to eliminate or even replace one or more postulates by other non-equivalent sentences in order to get non-Euclidian or non-Paschian geometries. Analogously it is possible to derive other set theories such that the axiom of choice, the continuum hypothesis, or Martin's axiom are no longer valid sentences. D. Hilbert considered that this sort of theoretical richness should be achieved in physics by means of a precise use of the axiomatic method.

	One important question in connection with the axiomatic method is to determine if a primitive concept of a given axiomatic system is definable or not on the basis of the remaining primitive concepts. For example: Usually force is a primitive concept in classical mechanics. But can we define ``force''? The answer depends on the axiomatic system that we employ in order to formulate classical mechanics (see, for example \cite{Sant'Anna-96}). In the next section we show how Padoa's principle may be used to prove the independence of primitive concepts.

\section{Padoa's Principle of Independence of Primitive Symbols}

	In an axiomatic system $S$ a primitive term or concept $c$ is definable by means of the remaining primitive ones if there is an appropriate formula, provable in the system, that fixes the meaning of $c$ in function of the other primitive terms of $S$. This formulation of definability is not rigorous but is enough here. When $c$ is not definable in $S$, it is said to be independent of the the other primitive terms.

	There is a method, introduced by A. Padoa \cite{Padoa-00}, which can be employed to show the independence of concepts. In fact, Padoa's method gives a necessary and sufficient condition for independence \cite{Beth-53,Suppes-57,Tarski-83}.

	In order to present Padoa's method, some preliminary remarks are necessary. Loosely speaking, if we are working in set theory, as our basic theory, an axiomatic system $S$ characterizes a species of mathematical structures in the sense of Bourbaki \cite{Bourbaki-68}. Actually there is a close relationship between Bourbaki's species of structures and Suppes predicates \cite{Suppes-67}; for details see \cite{daCosta-88}. On the other hand, if our underlying logic is higher-order logic (type theory), $S$ determines a usual higher-order structure \cite{Carnap-58}. In the first case, our language is the first order language of set theory, and, in the second, it is the language of (some) type theory. Tarski showed that Padoa's method is valid in the second case \cite{Tarski-83}, and Beth that it is applicable in the first \cite{Beth-53}.

	From the point of view of applications of the axiomatic method, for example in the foundations of physics, it is easier to assume that our mathematical systems and structures are contructed in set theory \cite{daCosta-88}.

	A simplified and sufficiently rigorous formulation of the method, adapted to our exposition, is described in the next paragraphs.

	Let $S$ be an axiomatic system whose primitive concepts are $c_1$, $c_2$, ..., $c_n$. One of these concepts, say $c_i$, is independent from the remaining if and only if there are two models of $S$ in which $c_1$, ..., $c_{i-1}$, $c_{i+1}$, ..., $c_n$ have the same interpretation, but the interpretations of $c_i$ in such models are different.

	Of course a model of $S$ is a set-theoretical structure in which all axioms of $S$ are true, according to the interpretation of its primitive terms \cite{Mendelson-97}.

	It is important to recall that, according to the theory of definition \cite{Suppes-57}, a definition should satisfy the {\em criterion of eliminability\/}. That means that a defined symbol should always be eliminable from any formula of the theory. 

	In the sequel we apply Padoa's method to some physical theories (i.e., axiomatic systems), in order to prove that time and space-time are eliminable (or dispensable) from some physical theories.

\section{McKinsey-Sugar-Suppes System of Particle Mechanics}

	This section is essentially based on the axiomatization of classical particle mechanics due to P. Suppes \cite{Suppes-57}, which is a variant of the formulation by J. C. C. McKinsey, A. C. Sugar and P. Suppes \cite{Suppes-53}. We call this McKinsey-Sugar-Suppes system of classical particle mechanics and abbreviate this terminology as MSS system.

	MSS system has six primitive notions: $P$, $T$, $m$, ${\bf s}$, ${\bf f}$, and ${\bf g}$. $P$ and $T$ are sets, $m$ is a real-valued unary function defined on $P$, ${\bf s}$ and ${\bf g}$ are vector-valued functions defined on the Cartesian product $P\times T$, and ${\bf f}$ is a vector-valued function defined on the Cartesian product $P\times P\times T$. Intuitivelly, $P$ corresponds to the set of particles and $T$ is to be physically interpreted as a set of real numbers measuring elapsed times (in terms of some unit of time, and measured from some origin of time). $m(p)$ is to be interpreted as the numerical value of the mass of $p\in P$. ${\bf s}_{p}(t)$, where $t\in T$, is a $3$-dimensional vector which is to be physically interpreted as the position of particle $p$ at instant $t$. ${\bf f}(p,q,t)$, where $p$, $q\in P$, corresponds to the internal force that particle $q$ exerts over $p$, at instant $t$. And finally, the function ${\bf g}(p,t)$ is to be understood as the external force acting on particle $p$ at instant $t$.

	Next we present the axiomatic formulation for MSS system:

\begin{definicao}
${\cal P} = \langle P,T,{\bf s},m,{\bf f},{\bf g}\rangle$ is a MSS system if and only if the following axioms are satisfied:

\begin{description}
\item [P1] $P$ is a non-empty, finite set.
\item [P2] $T$ is an interval of real numbers.
\item [P3] If $p\in P$ and $t\in T$, then ${\bf s}_{p}(t)$ is a
$3$-dimensional vector (${\bf s}_p(t)\in\Re^3$) such that $\frac{d^{2}{\bf s}_{p}(t)}{dt^{2}}$ exists.
\item [P4] If $p\in P$, then $m(p)$ is a positive real number.
\item [P5] If $p,q\in P$ and $t\in T$, then ${\bf f}(p,q,t) = -{\bf f}(q,p,t)$.
\item [P6] If $p,q\in P$ and $t\in T$, then $[{\bf s}_{p}(t), {\bf f}(p,q,t)] =
-[{\bf s}_{q}(t), {\bf f}(q,p,t)]$.
\item [P7] If $p,q\in P$ and $t\in T$, then
$m(p)\frac{d^{2}{\bf s}_{p}(t)}{dt^{2}} = \sum_{q\in P}{\bf f}(p,q,t) + {\bf g}(p,t).$
\end{description}
\end{definicao}

	The brackets [,] in axiom {\bf P6} denote external product.

	Axiom {\bf P5} corresponds to a weak version of Newton's Third Law: to every force there is always a counterforce. Axioms {\bf P6} and {\bf P5}, correspond to the strong version of Newton's Third Law. Axiom {\bf P6} establishes that the direction of force and counterforce is the direction of the line defined by the coordinates of particles $p$ and $q$.

	Axiom {\bf P7} corresponds to Newton's Second Law.

\begin{definicao}
Let ${\cal P} = \langle P,T,{\bf s},m,{\bf f},{\bf g}\rangle$ be a MSS system, let $P'$ be a non-empty subset of $P$, let ${\bf s}'$, ${\bf g}'$, and $m'$ be, respectively, the restrictions of functions ${\bf s}$, ${\bf g}$, and $m$ with their first arguments restricted to $P'$, and let ${\bf f}'$ be the restriction of ${\bf f}$ with its first two arguments restricted to $P'$. Then ${\cal P'} = \langle P',T,{\bf s}',m',{\bf f}',{\bf g}'\rangle$ is a subsystem of ${\cal P}$ if $\forall p,q\in P'$ and $\forall t\in T$, 
\begin{equation}
m'(p)\frac{d^{2}{\bf s}'_{p}(t)}{dt^{2}} = \sum_{q\in P'}{\bf f}'(p,q,t) + {\bf g}'(p,t).
\end{equation}
\label{P7}
\end{definicao}

\begin{teorema}
Every subsystem of a MSS system is again a MSS system.\footnote{In the original MSS system it is presented another definition for subsystem, where this theorem is not valid.}
\end{teorema}

\begin{definicao}
Two MSS systems \[{\cal P} = \langle P,T,{\bf s},m,{\bf f},{\bf g}\rangle\] and \[{\cal P'} = \langle P',T',{\bf s}',m',{\bf f}',{\bf g}'\rangle\] are equivalent if and only if $P=P'$, $T=T'$, ${\bf s}={\bf s}'$, and $m=m'$.
\end{definicao}

\begin{definicao}
A MSS system is isolated if and only if for every $p\in P$ and $t\in T$, ${\bf g}(p,t) = \langle 0,0,0\rangle$.
\end{definicao}

\begin{teorema}
If \[{\cal P} = \langle P,T,{\bf s},m,{\bf f},{\bf g}\rangle\] and \[{\cal P'} = \langle P',T',{\bf s}',m',{\bf f}',{\bf g}'\rangle\] are two equivalent systems of particle mechanics, then for every $p\in P$ and $t\in T$
\[\sum_{q\in P}{\bf f}(p,q,t) + {\bf g}(p,t) = \sum_{q\in P'}{\bf f}'(p,q,t) + {\bf g}'(p,t).\]\label{somaforcas}
\end{teorema}

	The embedding theorem is the following:

\begin{teorema}
Every MSS system is equivalent to a subsystem of an isolated system of particle mechanics.\label{Her}
\end{teorema}

	The next theorem can easily be proved by Padoa's method:

\begin{teorema}
Mass and internal force are each independent of the remaining primitive notions of MSS system.
\end{teorema}

	In the next paragraph follows a sketch of the proof of this last theorem.

	Consider, for example, two interpretations ($A$ and $B$) for MSS 
system, as it follows: In interpretation $A$ we have $P = \{1,2\}$, all  internal forces and external forces are null, $T$ is the real interval $[0,1]$, $s_p(t)$ is a constant $c$ for all $p$ and for all $t$, and $m_1 = m_2 = 1$; in interpretation $B$ we have $P = \{1,2\}$, all internal forces and external forces are null, $T$ is the real interval $[0,1]$, $s_p(t)$ is the same constant $c$ above for all $p$ and for all $t$, and $m_1 = m_2 = 2$. Since $s_p(t)$ is a given constant, then all accelerations are null. We 
can easily see that (i) these two interpretations are models of MSS (in the sense that they satisfy all the axioms of MSS); and (ii) these two models are the same, up to the interpretation of the masses $m_1$ and $m_2$. Then, according to Padoa's principle, mass is an independent concept and, so, is not definable. The same argument may used for internal force.

	According to Suppes \cite{Suppes-57}:

\begin{quote}
Some authors have proposed that we convert the second law [of Newton], that is, {\rm \bf P7}, into a definition of the total force acting on a particle. [...] It prohibits within the axiomatic framework any analysis of the internal and external forces acting on a particle. That is, if all notions of force are eliminated as primitive and {\rm \bf P7} is used as a definition, then the notions of internal and external force are not definable within the given axiomatic framework.
\end{quote}

	The next theorem is rather important for our discussion on the dependence of time with respect to the remaining primitive concepts.

\begin{teorema}\label{definet}
Time is definable from the remaining primitive concepts of MSS system.
\end{teorema}

\noindent
{\it Proof:} According to Padoa's Principle, the primitive concept $T$ in MSS system is independent from the remaining primitive concepts (mass, position, internal force, and external force) iff there are two models of MSS system such that $T$ has two interpretations and the remaining primitive symbols have the same interpretation. But these two interpretations are not possible, since position ${\bf s}$, internal force ${\bf f}$, and external force ${\bf g}$ are functions whose domains depend on $T$. If we change the interpretation of $T$, then we will change the interpretation of three other primitive concepts, namely, ${\bf s}$, ${\bf f}$, and ${\bf g}$. So, time is not independent and hence it can be defined.$\Box$\\

	The reader will note that our proof does not show how to define time. But that is not necessary if we want to show that time is dispensable in MSS system, since a definition does satisfy the criterion of eliminability.

\begin{teorema}\label{auto}
The differential equation given in axiom {\bf P7} is autonomous, i.e., it does not depend on an independent variable $t$ of time.
\end{teorema}

\noindent
{\it Proof:} Straightforward, since the parameter time $t$ is definable (theorem \ref{definet}) by means of position and forces.$\Box$\\

	What is the epistemological meaning of the definability of $T$? It is usual to say that one of the main goals of classical mechanics is to make predictions. But a prediction refers to the notion of future and, therefore, to the very notion of time. If time is dispensable, then one question remains: what is the main goal of classical particle mechanics, at least from the point of view of MSS system? It seems to us that the main goal of classical particle mechanics is to describe the physical state of particles. If we give differential equations\footnote{These differential equations will be necessarily autonomous, by means of theorem \ref{auto}.} (with respect to $t$) for ${\bf f}$ and ${\bf g}$, then the physical state of a particle $p$, in this framework, is the set of points

\begin{equation}
( {\bf s}_p(t), {\bf f}(p,q,t), {\bf g}_p(t))
\end{equation}

\noindent
in the phase space of position versus internal force versus external force, given initial conditions. By initial conditions we mean a specific point in the phase space. Since the phase space does not make any reference to the parameter $t$, then the set of points given above is equivalent to a set of points (a curve)

\begin{equation}
( {\bf s}_p, {\bf f}(p,q), {\bf g}_p)
\end{equation}

\noindent
given a specific point (initial condition) in the phase space.

\section{Other Physical Theories}

	One of us has worked with F. A. Doria in a series of papers concerning the axiomatic foundations of physical theories. Some of these papers are \cite{daCosta-91a,daCosta-91b,daCosta-92,daCosta-94}. In \cite{daCosta-92} there is a unified treatment, by means of axiomatic method, for the mathematical structures underlying Hamiltonian mechanics, Maxwell's electromagnetic theory, the Dirac electron, classical gauge fields, and general relativity. This unified treatment is given by the following definition.

\begin{definicao}\label{definefisica}
The species of structures (\`a la Bourbaki) of a {\em classical physical theory\/} is given by the 9-tuple

\[\Sigma = \langle M, G, P, {\cal F}, {\cal A}, {\cal I}, {\cal G}, B, \bigtriangledown\varphi = \iota\rangle\]

\noindent
where

\begin{enumerate}

\item $M$ is a finite-dimensional smooth real manifold endowed with a Riemannian metric (associated to space-time) and $G$ is a finite-dimensional Lie group (associated to transformations among coordinate systems). 

\item $P$ is a given principal fiber bundle $P(M,G)$ over $M$ with Lie group $G$. 

\item ${\cal F}$, ${\cal A}$, and ${\cal I}$ are cross-sections of bundles associated to $P(M,G)$, which correspond, respectively, to the field space, potential space, and current or source space.

\item ${\cal G}\subseteq \mbox{Diff}(M)\otimes {\cal G}'$ is the symmetry group, where Diff$(M)$ is the group of diffeomorphisms of $M$ and ${\cal G}$ is the group of gauge transformations of the principal fiber bundle $P(M,G)$.

\item $\bigtriangledown\varphi = \iota$ is a Dirac-like equation, where $\varphi\in {\cal F}$ is a field associated to its corresponding potential by means of a field equation and $\iota\in {\cal I}$ is a current. Such a differential equation is subject to boundary or initial conditions $B$.

\end{enumerate}

\end{definicao}

	We are obviously omitting some important details, which can be found in \cite{daCosta-92}. The unified picture of classical gauge field equations by means of Dirac equation may be found in \cite{Doria-86}. Our point, in this section, is to discuss the mathematical role of space-time in classical physical theories as presented in definition \ref{definefisica}.

\begin{teorema}
Space-time is dispensable in a classical physical theory, in the sense of definition (\ref{definefisica}).
\end{teorema}

\noindent
{\it Proof:} Padoa's Principle says that the primitive concept $M$ in $\Sigma$ is independent from the remaining primitive concepts iff there are two models of $\Sigma$ such that $M$ has two interpretations and all the other primitive symbols have the same interpretation. But these two interpretations are not possible, since all the remaining concepts, except $G$, depend on space-time $M$. Any change of interpretation related to $M$ will imply a change of interpretation of $P$, ${\cal F}$, ${\cal A}$, ${\cal I}$, ${\cal G}$, $B$, and $\bigtriangledown\varphi = \iota$. Therefore, space-time is not independent and hence it can be defined. So, according to the criterion of eliminability of definitions, space-time is dispensable.$\Box$\\

	As said in the Introduction, we intend to discuss the philosophical consequences of our results in future works. But one fact seems to be very clear here: since our definition of classical physical theory refers to classical field theories, this last theorem says that the concept of {\em field\/} is more fundamental than the notion of {\em space-time\/}. If this result is not satisfactory from the intuitive point of view, then our mathematical framework for classical field theories should be changed.

\section{Acknowledgments}

	We acknowledge with thanks some suggestions and criticisms made by Dr. Ulrich Mohrhoff and Dr. Mark Stuckey in connection with a previous version of this text. The idea of the definability of time in MSS system was originated from questions raised by two students (Humberto R. R. Quoirin and Tomas K. Breuckmann) of one of us (ASS) into the context of seminars on the logical foundations of physics at Federal University of Paran\'a.


\begin{thebibliography}{99}

\bibitem{Bender-85} Bender, C. M., K. A. Milton, D. H. Sharp, L. M. Simmons, Jr., R. Stong, `Discrete time quantum mechanics', {\em Phys. Rev. D\/} {\bf 32} 1476 (1985) 

\bibitem{Beth-53} Beth, E. W., `On Padoa's method in the theory of definition', {\em Indag. Math.\/} 330-339 (1953).

\bibitem{Bourbaki-68} Bourbaki, N., {\em Theory of Sets\/} (Hermann and Addison-Wesley, 1968).

\bibitem{Carnap-58} Carnap, R., {\em Introduction to Symbolic Logic and Its Applications\/} (Dover, New York, 1958).

\bibitem{daCosta-88} da Costa, N. C. A., and R. Chuaqui, `On Suppes' set theoretical predicates' {\em Erkenntnis\/} {\bf 29} 95--112 (1988).

\bibitem{daCosta-91a} da Costa, N. C. A.,  and F. A. Doria, `Undecidability and incompleteness in classical mechanics,' {\em Int. J. Theor. Phys.\/} {\bf 30}, 1041–-1073 (1991)

\bibitem{daCosta-91b} da Costa, N. C. A., and F. A. Doria, `Classical physics and Penrose's thesis', {\em Found. Phys. Lett.\/} {\bf 4}, 363–373 (1991)

\bibitem{daCosta-92} da Costa, N. C. A., and F. A. Doria, `Suppes predicates for classical physics', in J. Echeverria et al., eds., {\em The Space of Mathematics\/} (Walter de Gruyter, Berlin–New York, 1992).

\bibitem{daCosta-94} da Costa, N. C. A., and F. A. Doria, `Suppes predicates and the construction of unsolvable problems in the axiomatized sciences' in P. Humphreys, ed., {\em Patrick Suppes, Scientific
Philosopher, II\/}, 151-–191 (Kluwer, 1994).

\bibitem{Doria-86} Doria, F. A., S. M. Abrah\~ao and A. F. Furtado do Amaral, `A Dirac–like equation for gauge fields', {\em Prog. Theor. Phys.\/}, {\bf 75} 1440–1446 (1986).

\bibitem{Hilbert-00} Hilbert, D., `Mathematical problems', {\em Bull. Amer. Math. Soc.\/} {\bf 37} 407--436 (2000). Originally published as `Mathematische Probleme' {\em Vortrag, gehalten auf dem internationalen Mathematike-Congress zu Paris 1900\/}, G\"ott. Nachr., 253-297, (Vandenhoeck \& Ruprecht, G\"ottingen, 1900).

\bibitem{Norton-97a} Jaroszkiewicz, G. and K. Norton, `Principles of discrete time mechanics: I. Particle systems', {\em J. Phys. A\/} {\bf 30} 3115--3144 (1997).

\bibitem{Norton-97b} Jaroszkiewicz, G. and K. Norton, `Principles of discrete time mechanics: II. Classical field theory', {\em J. Phys. A\/} {\bf 30} 3145--3163 (1997).

\bibitem{Khorrami-95} Khorrami, M., `A general formulation of discrete-time quantum mechanics, restrictions on the action and the relation of unitarity to the existence theorem for initial-value problems', {\em Annals Phys.\/} {\bf 244} 101--111 (1995).

\bibitem{Lee-83} Lee, T. D., `Can time be a discrete dynamical variable?', {\em Phys. Lett. B\/} {\bf 122} 217 (1983).

\bibitem{Suppes-53} McKinsey, J. C. C., A. C. Sugar and P. Suppes, 
`Axiomatic foundations of classical particle mechanics', {\em J. Rational Mechanics and Analysis\/}, {\bf 2} 253--272 (1953).

\bibitem{Mendelson-97} Mendelson, E., {\em Introduction to Mathematical Logic\/} (Chapman \& Hall, London, 1997).

\bibitem{Mohrhoff-00} Mohrhoff, U., `What quantum mechanics is trying to tell us', {\em Am. J. Phys.\/} {\bf 68} 728--745 (2000).

\bibitem{Norton-98a} Norton, K. and G. Jaroszkiewicz, `Principles of discrete time mechanics: III. Quantum field theory', {\em J. Phys. A\/} {\bf 31} 977--1000 (1998).

\bibitem{Norton-98b} Norton, K. and G. Jaroszkiewicz, `Principles of discrete time mechanics: IV. The Dirac equation, particles and oscillons', {\em J. Phys. A\/} {\bf 31} 1001--1023 (1998).

\bibitem{Padoa-00} Padoa, A., `Essai d'une th\'eorie alg\'ebrique des nombres entiers, pr\'ec\'ed\'e d'une introduction logique \`a une th\'eorie d\'eductive quelconque'', {\em Biblioth\`eque du Congr\`es International de Philosophie}, {\bf 3} (1900).

\bibitem{Sant'Anna-96} Sant'Anna, A. S., `An axiomatic framework for classical particle mechanics without force', {\em Philosophia Naturalis} {\bf 33} 187-203 (1996).

\bibitem{Sant'Anna-99} Sant'Anna, A. S., `An axiomatic framework for classical particle mechanics without space-time', {\em Philosophia Naturalis\/} {\bf 36} 307--319 (1999).

\bibitem{Stewart-95} Stewart, I., {\em Nature's Numbers: The Unreal Reality of Mathematical Imagination\/} (Orion Pub. Group, 1995).

\bibitem{Stuckey-96} Stuckey, W. M., `Defining spacetime', {\em Astrophysics and Space Science\/} {\bf 244} 371-374 (1996).

\bibitem{Stuckey-99} Stuckey, W. M., `Leibniz's principle, dynamism, and nonlocality', {\em Physics Essays\/} {\bf 12} 414-419 (1999).

\bibitem{Stuckey-??} Stuckey, W. M., `Uniform spaces in the pregeometric modeling of quantum non-separability', forthcoming.

\bibitem{Suppes-57} Suppes, P., {\em Introduction to Logic\/}, (Van Nostrand, Princeton, 1957).

\bibitem{Suppes-67} Suppes, P., {\em Set-Theoretical Structures in Science\/}, mimeo. (Stanford University, 1967).

\bibitem{Tarski-83} Tarski, A., `Some methodological investigations on the definability of concepts', in A. Tarski, {\em Logic, Semantics, Metamathematics\/} 296--319 (1983).

\bibitem{Tati-64} Tati, T., `Concepts of space-time in
physical theories', {\em Prog. Theor. Phys. Suppl.} {\bf 29} 1--96 (1964).

\bibitem{Tati-83} Tati, T., `The theory of finite degree of
freedom', {\em Prog. Theor. Phys. Suppl.} {\bf 76} 186--223 (1983).

\bibitem{Vafa-00} Vafa, C., `On the future of mathematics/physics interaction', in V. Arnold, M. Atiyha, P. Lax, and B. Mazur (eds.) {\em Mathematics: Frontiers and Perspectives\/} (AMS, 2000).

\end{thebibliography}
\end{document}